\documentclass[aps,prd,twocolumn,groupedaddress,showpacs,graphicx]{revtex4}

\usepackage{amssymb,axodraw,graphics,graphicx}

\def\gs{\gtrsim}

\def\roughly#1{\mathrel{\raise.3ex\hbox
{$#1$\kern-.75em\lower1ex\hbox{$\sim$}}}}

\def\gs{\roughly>}
\newcommand{\pslash}{D\kern-0.15em\raise0.17ex\llap{/}\kern0.15em\relax}
\begin{document}

\title{Neutrino masses, muon g-2, dark matter, lithium problem, and leptogenesis at TeV-scale}
\author{Chian-Shu~Chen}\email{chianshu@phys.sinica.edu.tw}
  \affiliation{National Center for
Theoretical Sciences (South), Tainan, Taiwan 701}
  \affiliation{Institute of Physics, Academia Sinica, Taipei, Taiwan 115}
\author{Chung-Hsien Chou}\email{chouch@mail.ncku.edu.tw}
\affiliation{Department of Physics, National Cheng Kung
University, Tainan, Taiwan 701} \affiliation{National Center for
Theoretical Sciences (South), Tainan, Taiwan 701}
\date{Draft \today}
\begin{abstract}
Observational evidences of nonzero neutrino masses and the
existence of dark matter request physics beyond standard model. A
model with extra scalars and leptonic vector-like fermions is
introduced. By imposing a $Z_2$ symmetry, the neutrino masses as
well as anomalous muon magnetic moment can be generated via
one-loop effects at TeV-scale. An effort of explaining dark
matter, Lithium problem, and leptogenesis is presented. This
scenario can be tested at LHC and/or future experiments.
\end{abstract}

\maketitle
\section{Introduction}
There are some solid evidences for the physics beyond the standard
model (SM) of particle physics. One is the observations of
neutrino oscillations which has established that neutrinos have
very small masses. The low energy accelerator experiment of the
muon anomalous magnetic moment also gives another hint for the
physics beyond the SM. Besides that there are also evidences from
early Universe cosmology and astronomy: existence of dark matter
and matter-antimatter asymmetry of the Universe\cite{eta}. The
observed baryon asymmetry of the Universe can not be explained
within the SM with one CP violating phase.

Standard big-bang nucleosynthesis (SBBN) is one of the most
reliable and farthest reaching probes of early Universe cosmology.
One can calculate the relative abundances of light elements to H
at the end of the "first three minutes" after the big bang.
Despite the great success of SBBN, it has been noted that the
prediction for the ratio of $^7Li/H$ and the isotopic ratio
$^6Li/^7Li$ do not agree with current observations, called the
"lithium problems"\cite{PDG}. The SBBN model predicts primordial
$^6Li$ abundance about three orders of magnitude smaller than the
observed abundance level and $^7Li$ abundance a factor of two to
three larger when one adopts a value of the energy density of
baryon inferred from the WMAP data. They do not have an
astrophysical solution in a complete manner at present. One of the
plausible solution is the existence of primordial late-decaying charged
particles in the early Universe.

In this paper, we consider a novel model which can address all
these issues within a single framework. Our model is an extension
of the radiative seesaw mechanism of neutrino mass with extra
scalars and leptonic vector-like fermions. By imposing a $Z_2$
symmetry, the neutrino masses as well as anomalous muon magnetic
moment can be generated via one-loop effects at TeV-scale.

\section{The Model }
Besides the inert doublet model, we introduced a set of new
fermionic lepton doublet
 $L_i$ in our model. The new fermions are assumed to be vector-like to make sure that the theory is
 anomaly free as for self consistency. A discrete symmetry is imposed such that all the new
 particles are odd and the SM particles are even under this $Z_2$ projection. The content of the model
 is as following, scalar sectors
\begin{eqnarray}
\phi_{i =1,2} \quad {\rm and} \quad S^+
\end{eqnarray}
and an extra fermionic part
\begin{eqnarray}
L_i = \left(\begin{array}{c}N \\E^-\end{array}\right)_i,
\end{eqnarray}
where $\phi_1$ corresponds to the SM Higgs which is even under
$Z_2$. So we have the new Yukawa couplings
\begin{eqnarray}
L_Y &=& f_{\alpha i}\bar{l^c}_{\alpha}L_{i}S^+ + y_{\alpha i}l_{R\alpha}L_i\phi_2 + h.c. \nonumber \\
&=& \left[ f_{\alpha i}(\bar{\nu}_{\alpha}E^-_i +
l_{\alpha}^-\bar{N^c_i}) \right]S^+ \nonumber \\ &+& y_{\alpha
i}\left[ l^-_{R\alpha}E^+_i\phi^0_2 + l^-_{R\alpha}N_i\phi_2^-
\right] + h.c. ,
\end{eqnarray}
where $\alpha$ runs for e, $\mu$, and $\tau$, while $i$ stands for
the number of new fermionic doublet, we need at least two of them
in order to achieve successful leptogenesis.

The scalar potential is given by
\begin{eqnarray}
V(\phi_1,\phi_2,S^-) &=& -\mu_1^2|\phi_1|^2 + \lambda_1|\phi_1|^4
+ m_2^2|\phi_2|^2 + \lambda_2|\phi_2|^4 \nonumber \\ &+&
\lambda_3|\phi_1|^2|\phi_2|^2 + \lambda_4|\phi_1^{\dag}\phi_2|^2
\nonumber \\ &+& \frac{\lambda_5}{2}\left[ (\phi_1^{\dag}\phi_2)^2
+ h.c. \right] + m_s^2|S|^2 + \lambda_s|S|^4 \nonumber \\&+&
\mu\left[ (\phi_1^{0*}\phi_2^- - \phi_1^-\phi_2^0)S^+ + h.c.
\right].
\end{eqnarray}

Because the $Z_2$ symmetry is exactly conserved, $\phi_2^0$ can
not get VEV, and the symmetry breaking pattern is through
$\phi_1^0$ which can be identified as the SM Higgs $h$. The term
involving $\mu$ in the potential is interesting since it mix the
two new charged scalar, it is the important parameter associated
with neutrino mass matrix. The mixing matrix between $S^{\pm}$ and
$\phi_2^{\pm}$ is
\begin{eqnarray}
\left(\begin{array}{cc}\phi_2^+ &
S^+\end{array}\right)\left(\begin{array}{cc} m_2^2 +
\frac{\lambda_3v^2}{2} & \frac{\mu v}{\sqrt{2}} \\\frac{\mu
v}{\sqrt{2}} &
m_s^2\end{array}\right)\left(\begin{array}{c}\phi_2^-
 \\S^-\end{array}\right)
\end{eqnarray}

\subsection{Neutrino mass generation}
The neutrino masses can be generated at one-loop level  as shown
in Fig. 1. Note that there is a mixing between two charged scalars
$S^{\pm}$ and $\phi_2^{\pm}$ in the loop which
 associated with a GIM cancellation that make the corrections finite.

\begin{figure}[ht]
  \centering
    \includegraphics[width=0.25\textwidth]{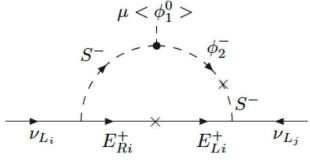}
  \caption{1-loop diagram for neutrino mass. }
  \label{fig:1loop}
\end{figure}

The generated neutrino mass matrix is
\begin{eqnarray}
(m_{\nu})_{\alpha\beta} &=& -i f_{\alpha i}f_{i\beta}M_{E_i}\mu^2<\phi^0_1>^2  \nonumber \\
&\times& \int{\frac{d^4q}{(2\pi)^4}}\frac{1}{(q^2 - M_s^2)^2}\frac{1}{(q^2 - M^2_{\phi_2})}
\frac{1}{(q^2 - M^2_{E_i})} \nonumber \\
&=& \frac{f_{\alpha i}f_{i\beta}\mu^2v^2M_{E_i}}{32\pi^2(M^2_{E_i}
- M^2_{\phi_2})}\left[ F(M^2_{E_i}) - F(M^2_{\phi_2}) \right]
\end{eqnarray}
where $F(M^2) = \frac{1}{(M^2 - M^2_s)} + \frac{M^2}{(M^2 -
M^2_s)^2}\ln{\frac{M^2_s}{M^2}}$.

Under the  assumption that $M_{E_i} \gs M_s \gs M_{\phi^-_2}$, the
neutrino masses can be approximated as
\begin{eqnarray}
(m_{\nu})_{\alpha\beta} &\approx& \frac{f_{\alpha
i}f_{i\beta}}{64\pi^2}\frac{\mu^2v^2}{M_{E_i}M^2_{\phi_2}} =
\frac{f_{\alpha i}f_{i\beta}}{64\pi^2}
(\frac{v}{M_{E_i}})(\frac{\mu^2}{M^2_{\phi_2}})v  \nonumber \\
 & \sim &  10^{-3}\times f^2\frac{\mu^2}{M_{E_i}}.
\end{eqnarray}
This mass matrix contains both a loop suppression factor and a
mass suppression factor hence it has similar structure as the
radiative seesaw models \cite{EMA}. Thus we obtain the results that
if $\mu$ is around $O(1) \sim O(100) GeV$, $f \sim 10^{-2}
- 10^{-4}$ to have the neutrino masses as $0.1 eV$, where we
already set $M_{\phi_2} \sim 500$ GeV which is from the constraint
of dark matter relic abundance \cite{DM}.

\subsection{Muon magnetic moment}
\begin{figure}[ht]
  \centering
    \includegraphics[width=0.5\textwidth]{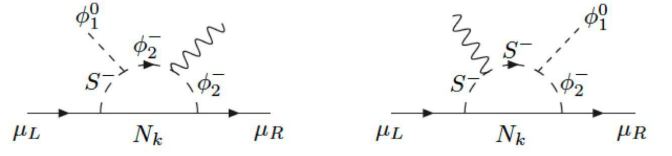}
  \caption{muon g-2 contributions from singly charged scalars mixing. }
  \label{fig:g-2_1}
\end{figure}

\begin{figure}[ht]
  \centering
    \includegraphics[width=0.3\textwidth]{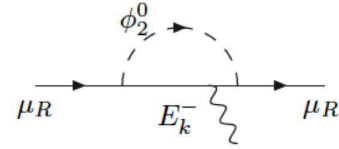}
  \caption{muon g-2 contributions from heavy charged leptons. }
  \label{fig:g-2_2}
\end{figure}

The current limit of the muon magnetic moment is \cite{PDG}
\begin{eqnarray}
\Delta a_{\mu} = (290\pm90)\times10^{-11},
\end{eqnarray}
which is $3.2\sigma$ deviation between theory and experiment. The
contributions to muon $g-2$ from our model are shown in Fig. 2 and
Fig. 3. In Fig. 2 one can see that the neutrino masses and the
muon $g-2$ are generated by similar mechanism. And we can get the
enhancement by the chirality flip in the internal fermion line,
the contributions are \cite{g-2}
\begin{eqnarray}
\Delta a^{NP}_{\mu(N_k)} =
-\frac{\sin{\delta}\cos{\delta}}{16\pi^2}\sum_k(f_{\mu k}y_{\mu
k}) \frac{m_{\mu}}{M_k}\left[ F(x_{P_1}) - F(x_{P_2}) \right],
\nonumber
\end{eqnarray}
where $x_{P_i} = m^2_{P_i}/M^2_k$ and $P_i$ are the mass
eigenstates of the charged scalars. And the function F is $F(x) =
\frac{1}{(1 - x)^3}[1 - x^2 + 2x\ln{x}]$. The mixing angle
satisfies the relation,
\begin{eqnarray}
\sin{\delta}\cos{\delta} = \frac{\mu v}{\sqrt{2}(m^2_{P_1} -
m^2_{P_2})},
\end{eqnarray}
this factor appears implicitly in neutrino masses  due to the GIM
mechanism. We found $\sin{\delta}\cos{\delta}\times f_{\mu k} \sim
(10^{-3} \sim 10^{-4})$, $m_{\mu}/(16\pi^2M_k) \sim 10^{-5}$, and
by setting $y_{\mu k} \sim O(10^{-1} - 10^{-2})$ will give us
magnetic moment of order $(10^{-9} \sim 10^{-10})$ which could
give enough contribution for the observed deviation between the SM
prediction and experiment.

For the contributions from the heavy charged leptons as shown in
Fig. 3, we see the enhanced FCNC type leading contribution in the
parameter regime where $m_{\mu} \ll M_{E^-_k},M_{\phi_2^0}$ and
$M_{E^-_k} \sim M_{\phi_2^0}$ is
\begin{eqnarray}
\Delta a^{NP}_{\mu(E^-_k)} &\approx& \frac{y_{\mu
k}^2}{48\pi^2}\frac{m^2_{\mu}}{M^2_{\phi_2^0}}\approx  y^2_{\mu
k}\times10^{-11}.
 \end{eqnarray}
It will not give us sufficient muon anomalous magnetic moment
unless the couplings $y_{\mu k}$ is of order of $O(10)$. We will
not consider this case in this paper.

\subsection{Dark matter, lithium problem and leptogenesis}
\underline{\emph{Dark matter}}. The neutral component of inert
doublet (ID) can be the dark matter candidate has been
investigated in \cite{DM}. The mass difference between the scalar
$(\phi^0_{2R})$ and pseudoscalar $(\phi^0_{2I})$  of the neutral
component is determined by the quartic coupling constant
$\lambda_5$ in the potential. To realize the dark matter relic
abundance, (co)annihilations of ID into gauge bosons or Higgs
should be carefully treated. The quartic couplings $\lambda's$ in
the potential are bounded by considering (co)annihilations
into/through Higgs. We should point out that besides the quartic
couplings $\lambda's$, the terms associated with $\mu$ in our model will also
contribute to the coannihilations between $S^-$ and $\phi_2^-$.
$\mu$ and $\lambda's$ can be related as $\mu \sim \lambda v$, thus
as similar to the discussions in \cite{DM}, $M_s - M_{\phi_2^-}$
can be constrained in a few GeV and $\mu$ is around $O(100)$ GeV.

\underline{\emph{Lithium problem}}. The lithium problem arises from the significant discrepancy
between the primordial $^7Li$ abundance as predicted by Standard
Big Bang Nucleosynthesis (SBBN) and the WMAP baryon density, and
the pre-Galactic lithium abundance inferred from observations of
metal-poor stars\cite{PDG,update}.

One of the solution to this is that the so-called Catalytic Big Bang
Nucleosynthesis (CBBN) \cite{lithium} which states if a long-lived
negatively-charged particle exists, it would form an exotic atom
and work as a catalyzer. The bound state will
 induce reactions which can produce suitable primordial abundance of
$^6Li$ and $^7Li$. In our model the scalar particle $S^-$ will
form the bound state with $^4He$ and this bound state will play
the role as the catalyzer. The catalytic path to $^6Li$ and $^9Be$
is
\begin{eqnarray}
&&S^- \rightarrow (^4HeS^-) \rightarrow ^6Li \quad {\rm and}
\nonumber \\  &&S^- \rightarrow (^4HeS^-) \rightarrow (^8BeS^-)
\rightarrow ^9Be.
\end{eqnarray}
And the key for the nuclear catalysis is an enormous enhancement
of the reaction rates in the photonless recoil reactions mediated
by $S^-$ :
\begin{eqnarray}
&& (^4HeS^-) + D \rightarrow ^6Li + S^- \quad {\rm and}  \nonumber \\
&& (^8BeS^-) + n \rightarrow ^9Be + S^-.
\end{eqnarray}

\begin{figure}[ht]
  \centering
    \includegraphics[width=0.5\textwidth]{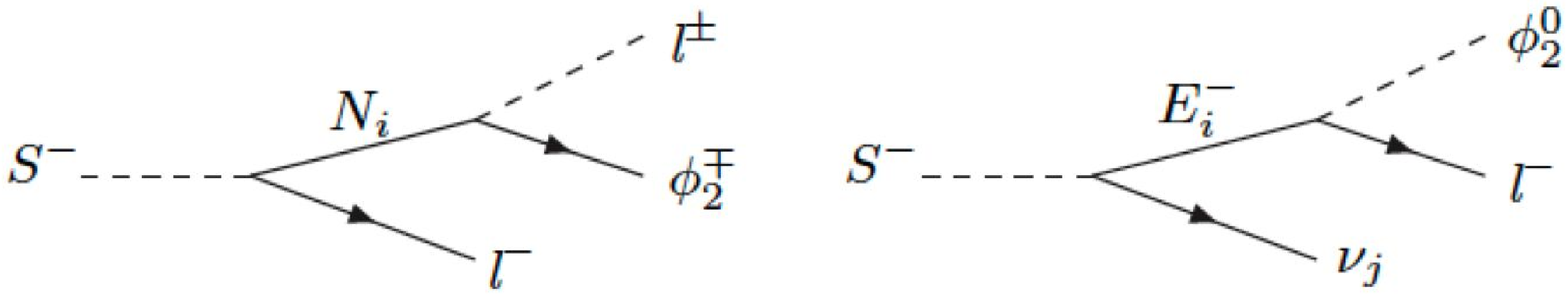}
  \caption{three body decays of $S^-$. }
  \label{fig:lithium}
\end{figure}

The rates of these catalyzed reactions depend sensitively on  the
abundance of $S^-$ at the relevant times. The observations impose
strong constraints on the lifetime of the negative charged
particle to be $\sim 10^3 sec$ to live long enough to form the
exotic atom and catalyze the reactions. In our model a long-lived
$S^-$ can be achieved through the three body decays into the
lepton sectors and dark matter in the final states as showed in
Fig. 4. With the new heavy leptonic doublet in the intermediate
states plus the small Yukawa couplings and the phase space
suppression of the mass differences between $S^-$ and $\phi_2$,
the long-lived $S^-$ can be easily realized. The decay rate of
$S^-$ is
\begin{eqnarray}
\Gamma_s|_{\alpha\beta(N_i)} &\approx& \frac{(f_{\alpha
i}y_{i\beta})^2}{30\pi^3M^4_{N_i}}
\times(\delta m)^5(1 - \frac{5m^2_l}{\delta m^2}) \nonumber \\
&\approx& f_{\alpha i}^2y_{i\beta}^2\times10^{-15}(\frac{\delta
m}{1GeV})^5 {\rm GeV},
\end{eqnarray}
where $\delta m = M_s - M_{\phi_2}$. The lifetime will
be around
\begin{eqnarray}
\tau_{\alpha\beta} \approx 6.6\times f^{-2}_{\alpha
i}y^{-2}_{i\beta}\times(\frac{\delta m}{1 GeV})^{-5}\times10^{-10} sec.
\end{eqnarray}
Combining the parameters from neutrino mass, muon magnetic moment,
and dark matter, i.e. $f \sim O(10^{-4} \sim 10^{-5})$ if $\mu
\sim O(100) GeV$, $y \sim (10^{-1} \sim 10^{-2})$, and $\delta m
\sim O(1) GeV$, one naturally obtains the lifetime
$\tau_{\alpha\beta}$ to be within the required range to solve the
lithium problem. Note that since the $M_s - M_{\phi_2^-} \sim O(1)
GeV$ from dark matter relic abundance, the three-body decay of
$S^-$ with a final $\tau$ lepton is kinematically suppressed.

\underline{\emph{Leptogenesis}}. There are two sources of CP
asymmetry  from each of the new Yukawa interactions as drawn in
Fig. 5. The decay rates of $N_1$ are the sum of
\begin{eqnarray}
\Gamma_{N_1} = \frac{\sum_{\alpha}(y_{1\alpha})^2}{16\pi}M_{N_1}
\quad {\rm and} \quad \Gamma_{N_1} =
\frac{(f^{\dag}f)_{11}}{8\pi}M_{N_1},
\end{eqnarray}

\begin{figure}[ht]
  \centering
    \includegraphics[width=0.5\textwidth]{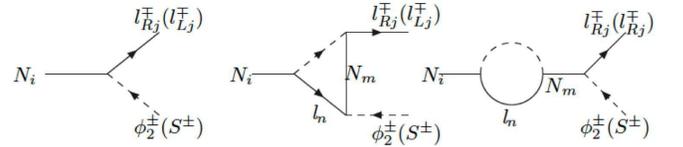}
  \caption{right-handed (left-handed) sector leptogenesis. }
  \label{fig:leptogenesis}
\end{figure}

which correspond to right-handed (left-handed) sector leptogenesis
in Fig. 5 respectively. The latter one is similar to the one in
canonical leptogenesis where its couplings are related to neutrino
masses. Let's consider the right-handed leptogenesis first
\cite{RH,RL}, we obtain the CP asymmetry
\begin{eqnarray}
\epsilon_1 &=& \frac{\Gamma(N_1 \rightarrow \emph{l} \phi_2^+) -
\Gamma(N_1 \rightarrow \bar{\emph{l}} \phi_2^-)}
{\Gamma(N_1 \rightarrow \emph{l} \phi_2^+) + \Gamma(N_1 \rightarrow \bar{\emph{l}} \phi_2^-)}  \nonumber \\
&=& \frac{1}{8\pi}\sum_{m\neq1}\frac{Im[(y^{\dag}y)^2_{1m}]}{\sum_{\alpha}(y^{\dag}y)_{1\alpha}}
\{ f_v(\frac{M^2_m}{M_1^2})
+ f_s(\frac{M_m^2}{M_1^2}) \} \nonumber \\
&=&
\frac{3}{16\pi}\sum_{m\neq1}\frac{Im[(y^{\dag}y)^2_{1m}]}{\sum_{\alpha}(y^{\dag}y)_{1\alpha}}\frac{M_1}{M_m},
\end{eqnarray}
where we have assumed the hierarchical masses of heavy neutrinos.
In this channel, the usual constraints from neutrino masses
disappear such that a hierarchical Yukawa couplings $y's$ can
easily be satisfied  to realize the amount of the baryon asymmetry
through leptogenesis, $\frac{n_B}{s} = -
\frac{28}{79}\frac{n_L}{s} = -1.36\times10^{-3}\epsilon_1\eta =
9\times10^{-11}$. For maximal efficiency, $\eta = 1$, we have the
relation \cite{RL}
\begin{eqnarray}
\frac{y^{(1)}}{y^{(2)}} <
0.28\times\sqrt{\frac{M_{N_1}}{M_{N_2}}\frac{M_{N_1}}{10^9GeV}}.
\end{eqnarray}
Note that $y^{(1)} = \sqrt{\sum_i|y_{1i}|^2} <
3\times10^{-4}(\frac{M_{N_1}}{10^9GeV})^{1/2}$ is bounded by the
out-of-equilibrium condition, and $y^{(2)} =
(\frac{Im[(y_{1\alpha})(y^*_{2\alpha})]^2}{\sum_{\alpha}(y_{1\alpha})(y^*_{1\alpha})})^{1/2}
\ge 1.05\times10^{-3}(\frac{M_{N_2}}{M_{N_1}})^{1/2}$ is the
amount of CP asymmetry we need. From these conditions, we can find
the TeV solution of leptogenesis, for example, if $M_{N_1} = 1
TeV$, $M_{N_2} = 5 TeV$, $y^{(2)} \simeq2.3\times10^{-3}$, and
$y^{(1)} \simeq3\times10^{-7}$. The effect of CP asymmetry from
the left-handed sector leptogenesis is constrained by the scale of
the neutrino mass and we find that the right-handed one gives the
dominate contribution.

Note that there will be extra washout effects due to gauge
interactions because of the non-trivial quantum numbers carried by
$N_i$. A similar washout effects can be found in type II seesaw
with decaying scalar triplet \cite{ST} and type III seesaw with
decaying fermionic triplet \cite{GaugeWashout}. It was shown in
\cite{ST,GaugeWashout} that due to the Boltzmann suppression
factor at temperatures below the gauge boson masses,  they can not
wash-out the lepton asymmetry in an efficient way. We should
emphasis here that the DM $\phi_2^0$ is formed in the decays of
$S^-$ (in the process of nucleosynthesis) and $N_i$ (in the
process of leptogenesis).

\subsection{Direct detection and Collider phenomenology}

Direct detection of DM can be measured through the elastic
scattering of a DM particle with a nuclei inside the detector. The
Z boson exchange channel constrains the lower bound of the mass
splitting $(M_{\phi^0_{2R}}-M_{\phi^0_{2I}})$  of order a few 100
keV\cite{DMZ}. The cross-section of the processes through exchange
a Higgs scalar $h$ at tree level and gauge bosons at one-loop
level are
 $\sigma^h \approx f^2_N\lambda_{\phi_2^0}^2/4\pi\times(\frac{m_N^2}{m_{DM}m^2_h})^2$ and
$\sigma_{1-loop} =
\frac{9f_N^2\pi\alpha_2^4m^4_N}{64M^2_W}(\frac{1}{M^2_W} +
\frac{1}{m^2_h})^2$. The later contribution is very interesting
because it is independent of DM mass which sets a lower bound
around $10^{-10}$ pb \cite{DM}. Although it is beyond the current
experimental sensitivity but it  is reachable in next-generation
experiments.

The new particles in our model are reachable at LHC. The
productions are dominated by the processes $q + \bar{q}
\rightarrow Z^* \rightarrow X + \bar{X}$, where $X$ represents
$S^-$, $\phi_2$, and $L_i$, and $q + q' \rightarrow W^*
\rightarrow \phi^-_2 (E_i^-) + \phi_2^0 (N_i)$. Note the
 new particles must be produced in pairs according to $Z_2$
symmetry. The novel signatures will be the missing energy while
producing the DM $\phi_2^0$ , and the decays of $\phi_2^-$ and
$E^-_i$ are mainly into $\phi_2^0\pi^-$ and $N_i\pi^-$
respectively. The charged particles in our model will leave
charged tracks in  detectors and  provide  clean signatures due to
their long lifetimes. One should note that $S^-$ and $\phi_2^+$
can only be produced through $h$ which is proportional to $\mu$.

\section{Summary and Discussions}

In summary, we propose a  model which can give neutrino mass with
radiative seesaw mechanism and can account for the muon anomalous
magnetic moment. In these two cases the GIM mechanism associated
with the scale parameter $\mu$ plays an important role. By setting
$\mu \sim O(100) GeV$ the relic abundance of dark matter from
inert doublet scalar can be realized, and a long-lived negative
charged particle which can fulfill the CBBN scheme for the Lithium
problem. The model also contains the low scale leptogenesis which
can help to solve the problem of baryon asymmetry in the Universe.
The nearly degenerate spectrum of the particle content will give
interesting collider phenomena  and can be tested in the near
future.

\begin{acknowledgments}
The research for this work has been supported in part by funds
from the National Science Council of Taiwan under Grant No.
NSC97-2112-M-006-004-MY3, and the National Center for Theoretical
Sciences, Taiwan.
\end{acknowledgments}

\end{document}